# In-flight detection of few electrons using a singlet-triplet spin qubit

Vivien Thiney,[1] Pierre-André Mortemousque,[1] Konstantinos Rogdakis,[1] Romain Thalineau,[1] Arne Ludwig,[2] Andreas D. Wieck,[2] Matias Urdampilleta,[1] Christopher Bäuerle,[1] and Tristan Meunier[1]

[1]*Université Grenoble Alpes, CNRS, Grenoble INP, Institut Néel, 38000 Grenoble, France*
[2]*Lehrstuhl für Angewandte Festkörperphysik, Ruhr-Universität Bochum, Universitätsstrasse 150, 44780 Bochum, Germany*

We investigate experimentally the capacitive coupling between a two-electron singlet-triplet spin qubit and flying electrons propagating in quantum Hall edge channels. After calibration of the spin qubit detector, we assess its charge sensibility and demonstrate experimentally the detection of less than five flying electrons with average measurement. This experiment demonstrates that the spin qubit is an ultrasensitive and fast charge detector with the perspective of a future single-shot-detection of a single flying electron. This work opens the route toward quantum electron optics experiments at the single-electron level in semiconductor circuits.

## I. INTRODUCTION

Constant progress in material science and nanofabrication has led to a rapid development of electronic circuits at the single-electron level [1] with the aim to perform quantum optics experiments with electrons rather than photons. The existing Coulomb coupling between electrons provides a mean for quantum manipulation hardly possible with photons [2]. To push the field of electron quantum optics [3] to the level of its photonic counterpart, a key ingredient—the single-shot electron detector—is still missing.

Using an on-chip nanometric electrometer, single-shot detection of an electron can presently only be achieved when the electron is static for a sufficiently long time [4–6]. Progress in detection efficiency has pushed down the acquisition time slightly below 1 $\mu$s. This has been demonstrated for electrons trapped in quantum dots [7–10], and is exploited for spin-based quantum information processing in semiconductors. Detecting an electron in flight, however, is much more challenging. In this case, the interaction time between a flying electron propagating at the Fermi velocity and an electrometer with an interaction radius of 1 $\mu$m [cf. Fig. 1(a)] is set by the width of the electron wave packet and is usually limited to 1 ns in semiconductor circuits [11]. Thus, an improvement of several orders of magnitude in charge sensitivity in comparison with the one demonstrated for trapped electrons is needed to enable in-flight detection of electrons.

Over the past decade, extraordinary results have been obtained with flying electrons in quantum interferometry [12–18] and one- and two-electron correlation experiments [19–23]. In the latter experiments, single-electron injection is periodically repeated and the DC current or the low-frequency current noise measurements provide information on the average value and the fluctuations of the charge that arrives in each contact. For electronic flying qubit experiments [1,2,24], on the other hand, it is compulsory to detect single flying electrons efficiently to allow for coincidence measurements and access high-order quantum correlations. This task being impossible with the best continuous electrometer demonstrated so far, we present a novel stroboscopic detector composed of such continuous electrometer combined with a quantum system. Indeed, quantum systems are extremely sensitive to environmental fluctuations (external perturbations) [25], and are well adapted to detect single quantum objects. For instance, they have already been used to detect a single phonon excitation of a nanomechanical system [26–28]. In AlGaAs two-dimensional electron gas (2DEG) systems, double quantum dot charge qubits or a Mach-Zehnder interferometry have been proposed to detect flying electrons [29].

Here we propose and experimentally demonstrate the in-flight electron detection by using a singlet-triplet spin qubit detector. The electrons propagate in the edge channels (ECs) of the quantum Hall effect (QHE) [30,31] as edge magnetoplasmons (EMPs). By capacitive coupling, they interact with the spin qubit detector defined in a double quantum dot. It results in a phase shift of the spin qubit. We extract a $\pi$ phase shift for $90 \pm 5$ flying electrons. When averaged, we can detect a signal from as low as 4 flying electrons. The presented results confirm the potential of such a charge detector for the single-shot detection of a single flying electron.

## II. SAMPLE AND SETUP

To implement the in-flight single-shot detector, we use GaAs/AlGaAs heterostructures where the electrons are propagating within the ECs [13,14] along the trajectories imposed by the surface gates as indicated by the red line in Fig. 1(a). A magnetic field of 430 mT is applied perpendicular to the surface to define the ECs at filling factor $\nu = 10$ [see

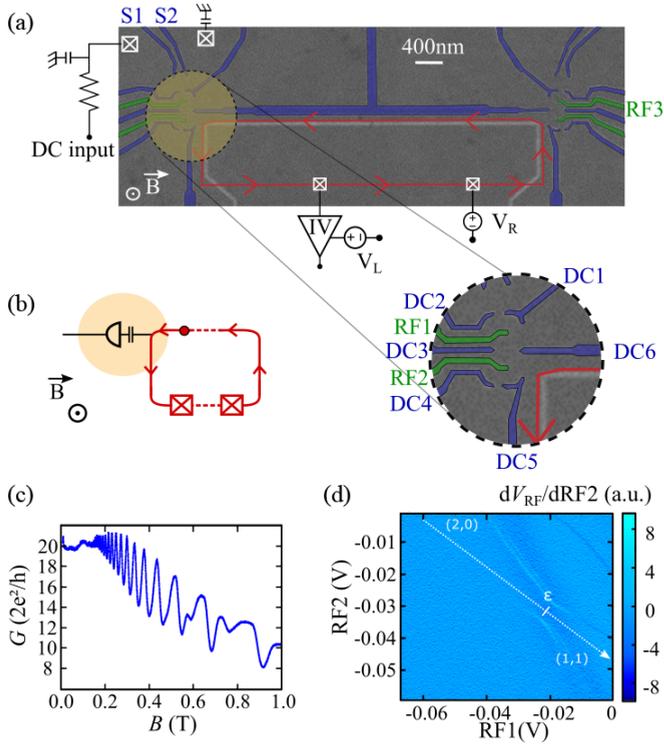

FIG. 1. (a) Scanning electron micrograph of the sample used. The DC gates are colorized in blue, the RF ones in green. The orange disk corresponds to the relevant interaction area capacitively coupled to the qubit. The blue line represents the 10 edge channels with their electron density controlled with $V_R$ ($V_L$) for the upper (lower) path. (b) Sketch of the sample with the spin qubit detector capacitively coupled to the ECs in which flying electrons illustrated with the dot are propagating. (c) Stability diagram using only the RF gates. The derivative of the RF sensing quantum dot signal with respect to $V_{RF2}$, $dV_{RF}/dV_{RF2}$, is plotted as a function of the swept RF2 gate voltage ($y$ axis) and stepped RF1 one ($x$ axis). This map is measured with 1 $\mu$s per point repeated 31 times. The indicated voltages are estimated values on the sample considering the AWG amplitude and the RF coaxial lines attenuations. The light blue lines indicate a change in the double quantum dot charge configuration labeled in white. For instance, (1,1) corresponds to the confinement of one electron in each quantum dot and (2,0) the two electrons in the quantum dot the farthest to the ECs. (d) Shubnikov–de Haas oscillations of the conductance $G$ as a function of the magnetic field $B$. The electron density of the 2DEG is evaluated to $n_e = 1.2 \times 10^{11}$ cm$^{-2}$. The mobility of the sample was independently measured equal to approximately 200 m$^2$ V$^{-1}$ s$^{-1}$. At a magnetic field of 430 mT, the QHE is not fully developed and the Shubnikov–de Haas oscillations start to show deviation from pure oscillating behavior.

Shubnikov–de Haas oscillations presented in Fig. 1(c)]. To probe the flying electrons, we engineer a two-electron spin qubit on the electrons propagation path. The spin qubit is obtained by confining two electrons in a double quantum dot defined with electrostatic surface gates (Ti-Au) deposited on the top of the heterostructure. DC and RF gates are used to control the spin qubit, shown respectively in blue and green colors in Fig. 1(a). Nanosecond control is possible by using RF gates excited with an arbitrary wave-form generator (AWG; Tektronix 5014). DC gates DC1 and DC5 [see Fig. 1(a) for labels] are used to control the coupling to the electron reservoir, tuned in the MHz regime. DC2 and DC4 control the quantum dots chemical potentials while gates DC3 and DC6 are used to tune the interdot tunnel coupling in the GHz regime.

The readout of the two-electron spin qubit is performed with the help of a sensing quantum dot (SQD) defined with gates S1 and S2, and embedded in a radio-frequency (RF) circuit [9]. To achieve optimal signal-to-noise ratio, the tank circuit is impedance-matched to the 50 $\Omega$ impedance of the transmission line at the maximal charge sensitivity of the SQD. Change in the charge occupancy of the double dot induces a variation of the SQD conductance ($g_{SQD}$) and leads to a change of the circuit impedance. It is measured with an amplitude demodulation technique and the readout circuit bandwidth is about 15 MHz. To demonstrate fast charge readout, we measure a charge stability diagram of the double quantum dot around the (2,0)-(1,1) charge transition, where ($N_1$, $N_2$) corresponds to the charge configuration with $N_1$ electrons confined in the quantum dot closest to the SQD and $N_2$ in the other dot. The results are shown in Fig. 1(d) where we set the acquisition time to 1 $\mu$s per point averaged 31 times. With our setup, because of the dependency of the tank circuit inductance to the magnetic field, we limit it to 430 mT to work with the charge sensitivity just shown.

## III. SPIN QUBIT AS CHARGE DETECTOR

The dynamic of two-electron spin qubits is highly dependent on the energy detuning $\varepsilon$ between the two dots and, as a consequence, they have been identified as good probes for charge detection [33]. Such a system is used as a very sensitive charge detector by implementing the coherent exchange oscillations between antiparallel spin states $|\uparrow\downarrow\rangle$ and $|\downarrow\uparrow\rangle$. These oscillations are the results of the exchange of a quantum of spin between the two adjacent electrons at a frequency proportional to the exchange energy $J(\varepsilon)$ [33–35]. As a consequence of the applied 430 mT magnetic field, the parallel spin states $|T_{-,+}\rangle$ are repelled away from the antiparallel ones to avoid spin mixing during coherent exchange oscillations. The spin qubit is first initialized in the singlet spin state $|S\rangle$ in the (2,0) region and transferred to the (1,1) region in the lowest energy antiparallel spin states using a combination of adiabatic and nonadiabatic passages [for details see Figs. 2(a) and 2(b)]. A rapid, nonadiabatic detuning pulse switches on the exchange interaction and induces coherent quantum oscillations between the antiparallel spin states. Reversing the complete pulse sequence to the (2,0) region maps the lowest spin state $|\uparrow\downarrow\rangle$ on the (2,0) charge configuration and $|\downarrow\uparrow\rangle$ on the (1,1) charge configuration. Since these two charge configurations are coupled differently to the sensing quantum dot, they result in two different RF-SQD voltages $V_{RF}$. The signal variations $\Delta V_{RF}$ of $V_{RF}$ are therefore directly proportional to the $|\uparrow\downarrow\rangle$ population.

Having introduced the experimental setup as well as the measurement scheme, we now discuss the charge sensitivity of the two-electron spin qubit to the electrostatic environment. We perform coherent exchange oscillations at different detuning positions $\varepsilon$ [35] and for different pulse durations following the sequence shown in Fig. 2(b). A typical two-dimensional map of the detector response $\Delta V_{RF}$ is presented in Fig. 2(c). These data are fitted to determine the exchange energy $J(\varepsilon)$ expression which is the following:

$$J(\varepsilon(t)) = J_0 + J_1 e^{\frac{\varepsilon(t)}{\sigma}}, \quad (1)$$

$$\varepsilon(t) = \left(1 - e^{-(\frac{t}{\tau})}\right)\varepsilon_{AWG} - \varepsilon_0, \quad (2)$$

with $J_0 = 10$ neV, $J_1 = 4$ μeV, $\sigma = 1.5$ mV, $\tau = 0.85$ ns, and $\varepsilon_0$ the reference position of the (1,1)-(2,0) charge transition.

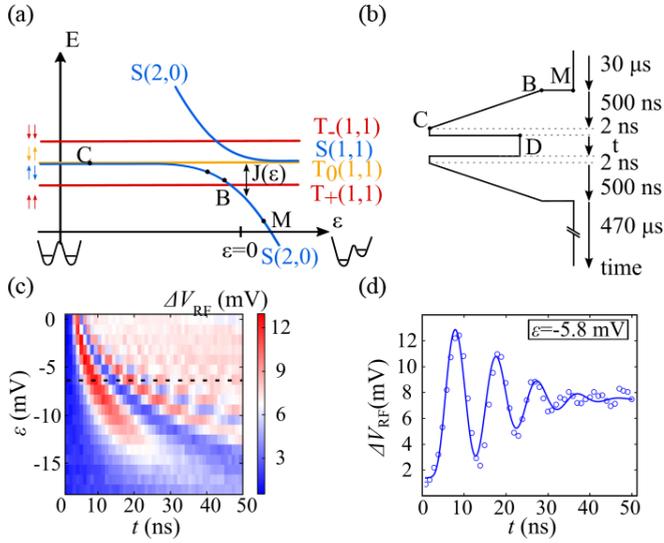
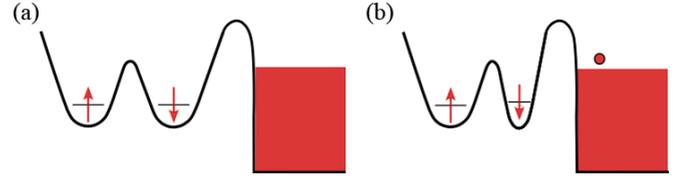

FIG. 2. (a) Energy diagram of the two electron spin states in a double quantum dot as a function of the detuning $\varepsilon$ for the considered charge transition (1,1)-(2,0); the detuning position $\varepsilon = 0$ is defined as the transition between these two charge states. The exchange energy $J(\varepsilon)$ is the energy splitting between the spin states $|S\rangle$ and $|T_0\rangle$ which varies with $\varepsilon$. The different letters are the different points of the sequence to induce coherent exchange oscillations. (b) Sketch of the pulse sequence with the different timings used in the experiment. (c) $\Delta V_{RF}$ as a function of the oscillating time ($x$ axis) and the oscillating position $\varepsilon$ ($y$ axis). Each point corresponds to an average over typically 5000 identical experimental sequences. The extracted $\Delta V_{RF}$ is plotted in color scale with in blue (red) the low (high) intensity. The map is characterized by the acceleration of the coherent exchange oscillations when increasing the detuning value $\varepsilon$. The black dashed line indicates the cut along which the curve shown in (d) is extracted. The low-reported oscillation contrast obtained for small $\varepsilon$ is due to the partial adiabaticity of the exchange pulse [32]. (d) Single trace of coherent exchange oscillations extracted at an equivalent oscillating position to the one indicated by the black dashed line of (c). The dashed lined is the result of the fit of the data $P_{|\uparrow\downarrow\rangle}(t) = \frac{1}{2}[1 - e^{-(t/T_2^*)^2} \cos(\int_0^t \frac{J(\varepsilon(t))dt}{\hbar})]$, with $T_2^*$ the dephasing time set to 20 ns [33,34].

A single trace for fixed detuning $\varepsilon$ (black dashed line) is shown in Fig. 2(d). One observes an exponential increase of the oscillation frequency when $\varepsilon$ approaches zero detuning. This phenomenon, already seen in previous works [32,35], is interpreted as a signature that the gate voltages applied to vary the detuning change also the dot positions. In the case of the exchange pulse in $\varepsilon$, it brings the two dots closer and changes also the tunneling process between the two dots. This experiment confirms the strong sensitivity of the coherent exchange oscillations to the electrostatic environment of the double quantum dot, materialized here by the energy detuning $\varepsilon$. It is this phenomenon which will be exploited for the detection of flying electron wave packets. An electron passing nearby the detector, as sketched in Fig. 3, induces a variation in energy detuning $\delta\varepsilon$ for a finite time and consequently to a phase shift of the coherent exchange oscillations. As a result, the passage of the flying electron will be mapped onto the population of the S-$T_0$ spin qubit. In comparison with other electrometers, the advantage of using two-electron spin qubits for charge detection is twofold: when tuned into the charge-sensitive regime, the spin qubit detector can be used as a large detection-bandwidth charge detector. Second, the information is encoded in the spin qubit population which can be stored for up to several hundred microseconds [36]. This time is long enough to allow single-shot readout with a conventional charge detector after spin-to-charge conversion. Ideally, an induced phase shift equal to $\pi$ for a given interaction time results in a complete spin exchange and hence unity detection efficiency. Thus, combining a $\pi$ phase shift and single-shot spin readout of the S-$T_0$ qubit, possible by employing a RF-SQD, provides an in-flight single-shot flying electron detection.

FIG. 3. Sketch of two electrons confined in a double quantum dot, without (a) and with (b) a flying electron in the nearby ECs. When a flying electron is passing in the vicinity of the qubit detector, the detuning is changed and the two confined electrons get closer by capacitive coupling modifying the qubit dynamics.

## IV. INFLUENCE OF A CONTINUOUS STREAM OF ELECTRONS ON THE CHARGE DETECTOR

To investigate the spin qubit response to flying electrons, i.e., local and propagating change of the ECs electron density, a continuous stream of electrons is generated by applying a DC bias voltage between $V_R$ and $V_L$ [see Figs. 1(a) and 4]. From the Landauer-Büttiker formalism [37,38], the chemical potential of the ECs $\mu_{\text{edge}}$ is defined by the contact which is emitting the electrons as depicted in Fig. 4. For a positive magnetic field, the electrons travel anticlockwise at a potential $\mu_{\text{edge}} = \mu_R$, without any dependence on $\mu_L$. With opposite magnetic field, the propagation direction is reversed with a potential fixed by $\mu_{\text{edge}} = \mu_L$. Coherent exchange oscillations while stepping the bias of the two ohmic contacts for positive and negative magnetic fields are presented in Fig. 5. A phase shift is only observed when stepping the bias of the contact controlling the electron density of the ECs passing nearby the double quantum dot. Such an observation is first a proof of the strong chirality at a micrometer scale even if the QHE is not fully developed over macroscopic distances (cf. Fig. 1). Second, it clearly demonstrates the sensitivity of the spin qubit detector to a change in the EC electron density. Here, a $\pi$ shift is observed for a bias difference of 600 $\mu$V and an interaction time of 30 ns. The effect of the flying electrons on the coherent exchange oscillations is similar to that of a polarized gate electrode. Since the ECs are capacitively coupled to the two electrons of the spin qubit detector, varying the bias induces a small detuning shift $\delta\varepsilon$, hence, a change in the oscillation frequency. More precisely, the exact position of the electrons forming the spin qubit is crucial to understand the detuning shift induced by the DC bias. Indeed, the singlet-triplet spin qubit is realized at the (1,1)-(2,0) charge transition, where the (2,0) charge configuration corresponds to a situation where the two electrons are located in the quantum dot that is the farthest away from the ECs. For positive (negative) bias, the ECs electron density is increased (decreased), the two electrons of the qubit are pushed farther away (attracted closer) to the (2,0) charge configuration, and the resulting exchange oscillations are slower (faster). In this experiment it was the global ECs electron density that was varying while, in the following, a local and propagating variation of it defined as EMPs will be used.

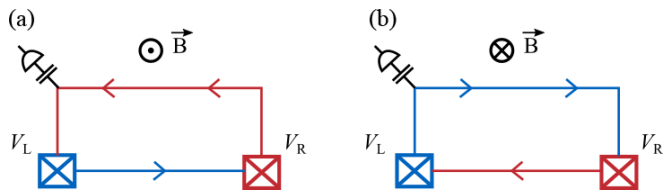

FIG. 4. (a) Sketch of the sample with the edge channels represented by a single line; the color indicates which ohmic contact set their electron density, blue for the left, red for the right. The edge channels passing nearby to the spin qubit detector, so capacitively coupled to it, have a chemical potential defined by the right ohmic contact, $\mu_{\text{edge}} = \mu_R = E_F - eV_R$, and is insensitive to changes of the potential applied to the left contact. (b) Same sketch but for an opposite magnetic field, then $\mu_{\text{edge}} = \mu_L = E_F - eV_L$.

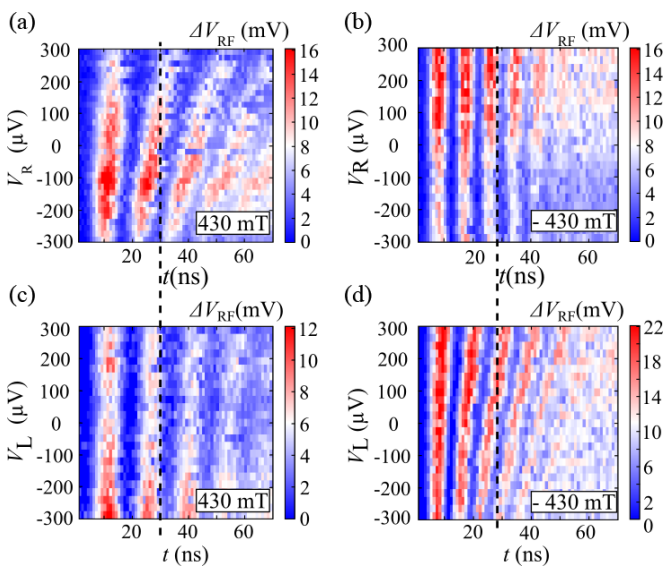

FIG. 5. (a) $\Delta V_{\text{RF}}$, the voltage amplitude of the singlet triplet RF-SQD signal, is plotted in color scale as a function of the exchange pulse duration, swept from 1 to 100 ns, along the x axis and the DC bias $V_R$ ($V_L$), stepped from $-300$ to $300$ $\mu$V, along the y axis. For these two measurements, the magnetic field was set to 430 mT. The dashed line is a guide for the eye to highlight the phase shift when $V_R$ is stepped. It remains constant when stepping $V_L$. The evolution of the $\Delta V_{\text{RF}}$ amplitude within a map is due to variation of the RF-SQD charge sensitivity in time. (b) Same measurements but for the reversed magnetic field. This time the phase shift is observed when stepping $V_L$. For these maps the coherent exchange oscillations are faster because of a small sample tuning variation when reversing the magnetic field. These data have been measured at a detuning of $-10.7$ mV with a different sample tuning, i.e., interdot tunnel coupling lower, than for the rest of measurements presented in this paper.

## V. SPIN QUBIT DETECTOR CALIBRATION

To investigate the potential of using such a spin qubit to detect individual flying electrons, it is important to understand the qubit dynamic at positive detuning where maximum detector sensitivity is expected. Indeed, the more positive is the detuning, the faster are the oscillations and the larger is the phase shift induced by the flying electrons. Our goal is to obtain the largest charge sensitivity of the detector to probe a flying electrons wave packet generated by a subnanosecond pulse presented in Fig. 8. Therefore, the exchange pulse of the spin qubit detector, the spin oscillating time, was set to its minimum value in our setup of 1 nanosecond for demonstrating the principle of the experiment. The strategy to optimize the detection was then to sweep the detuning in order to identify the best sensitivity point. The resulting coherent oscillations of the qubit population are presented in Fig. 6(a). They are the results of the accumulated phase during the exchange pulse. Such a phase is defined as $\int_0^t \frac{J(\varepsilon(t))dt}{\hbar}$ with $J(\varepsilon) \propto \exp(\varepsilon)$ leading to an increase of the oscillation frequency with detuning. These oscillations are fitted using an expression derived from the model used in Fig. 2(c). We use a Gaussian model for the pulse in $\varepsilon(t)$; we integrate the evolution between $\pm t_{int}$ with $t_{int} = 2.5$ ns and assume that most of the dephasing occurs close to the maximum of the pulse amplitude for a duration $\tau = 0.2$ ns:

$$P_{|\uparrow\downarrow\rangle}(\varepsilon) = \frac{1}{2}\left[1 - e^{-[\tau/T_2(\varepsilon)^*]^2} \cos\left(\int_{-t_{int}}^{+t_{int}} \frac{J(\varepsilon(t))dt}{\hbar}\right)\right], \quad (3)$$

$$\varepsilon(t) = \alpha e^{-(\frac{t}{\tau})^2}\varepsilon_{AWG} - \varepsilon_0, \quad (4)$$

$$T_2(\varepsilon)^* = \Gamma \hbar \frac{d\varepsilon}{dJ}, \quad (5)$$

where $\alpha = 0.84$ represents the reduction of the pulse amplitude extracted from the finite bandwidth of the signal generator and $\varepsilon_0$ the reference position of the (1,1)-(2,0) charge transition. In addition to the oscillation amplitude, there is a second fitting parameter $\Gamma = 4000$ V$^{-1}$, which is the inverse of the effective gate noise and is consistent with previous reported values [39]. Only two oscillations are visible indicating a rapid reduction of the coherence time with detuning. This is attributed to the increased sensitivity of the system to charge noise as the system is brought closer to the (1,1)-(2,0) crossing [35]. To determine the actual charge sensitivity of the spin qubit detector, a preliminary step is to determine its actual exposure time $\tau$. Indeed, because of the nonlinear evolution of the exchange energy $J(\varepsilon)$, the phase is mostly accumulated when the exchange pulse approaches its maximum amplitude. We define $\tau$ as the time required to accumulate 50% of the total phase (from 25% to 75%) and estimate it from the exchange pulse recorded at room temperature, fitted with a Gaussian expression then cut in time bins. For each bin, using the exchange energy expression obtained from experimental data [shown in Fig. 6(b)] we calculate the accumulated phase to obtain an exposure time about $\tau = 200$ ps.

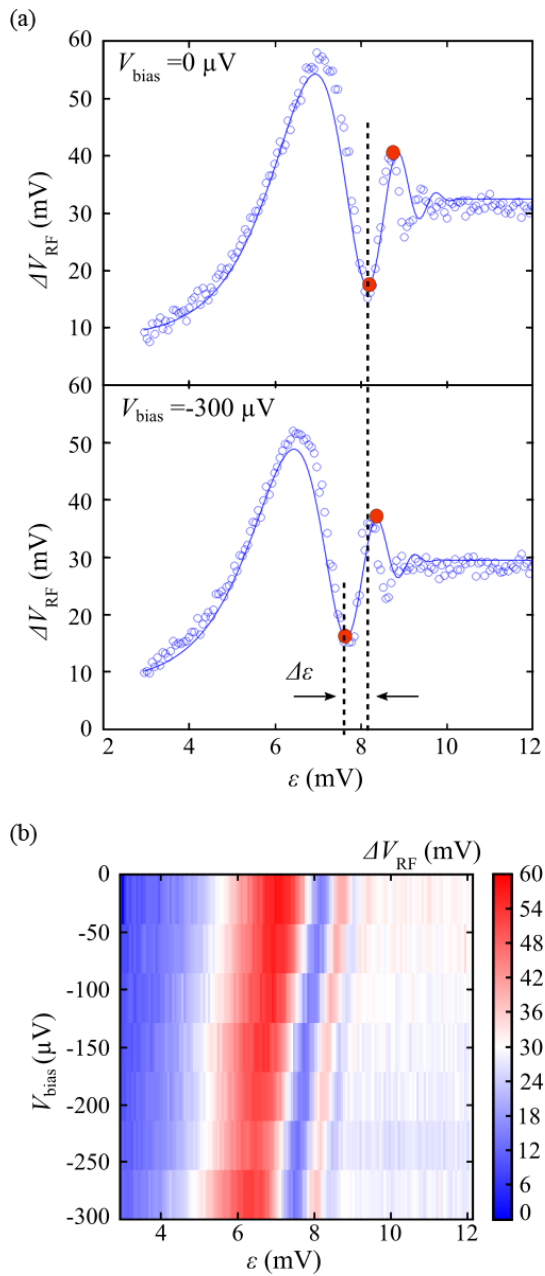

FIG. 6. (a) The voltage amplitude $\Delta V_{\text{RF}}$ of the spin triplet RF-SQD signal evolution is plotted as a function of the detuning $\varepsilon$. For such coherent exchange oscillations in amplitude the interaction time is set to 1 ns. The solid line is the result of the data fitting described in the main text. These traces are extracted from the map shown in (b) for $V_{\text{bias}} = 0$ and $V_{\text{bias}} = -300$ μV. The two red dots indicate the $2\pi$ and $3\pi$ spin rotation position. (b) Voltage amplitude $\Delta V_{\text{RF}}$ as a function of $\varepsilon$ for different DC bias applied to the ECs. By applying a negative $V_{\text{bias}}$, the EC density increases resulting in an acceleration of the oscillations, that corresponds to a linear shift in $\varepsilon$.

Similar to the coherent exchange oscillations in time (see Fig. 5), an acceleration of the oscillations while increasing the ECs electron density is observed in Fig. 6(b). For a DC bias of 300 μV, a complete $\pi$ phase shift is obtained, between the $2\pi$ and $3\pi$ spin rotation, corresponding to a change in the detuning pulse amplitude of $\Delta \varepsilon = 0.7$ mV. It is highlighted with the black dashed lines in Fig. 6(a). We observe a linear dependence between the detuning pulse amplitude and the edge state bias [see Fig. 6(b)], consistent with the interpretation that the ECs act as a gate for the spin qubit.

To obtain quantitatively the sensitivity in flying electron number, the voltage bias has to be converted in the number of electrons $N_e$ interacting with the spin qubit detector during the exposure time $\tau$. It is obtained via the relation $N_e = \frac{I\tau}{e}$, where $e$ is the elementary charge of the electron, $I$ the current flowing in the ECs, and $\tau$ the detector exposure time. The evolution of the detuning point $\varepsilon_{2\pi}$ that corresponds to the detuning where a $2\pi$ spin rotation has been accomplished is therefore analyzed with increasing $N_e$ in Fig. 7(a). We observe a linear shift of $\varepsilon_{2\pi}$ that permits us to evaluate the expected shift for a single flying electron $\delta\epsilon$, equal to 6 μV.

Finally, we estimate in Fig. 7(b) the expected detection RF signal $\delta V_{\text{RF}} = \delta V_{\text{RF}}(\varepsilon + \delta \varepsilon) - \delta V_{\text{RF}}(\varepsilon)$ induced by a single electron in our setup. We identify the maximum qubit detector charge sensitivity, i.e., the largest $\delta V_{\text{RF}}$, obtained for two different exchange pulse amplitudes $\varepsilon$, 7.5 mV and 8.5 mV, respectively the $3\pi/2$ and $5\pi/2$ spin rotation positions, with $\delta V_{\text{RF}} \approx 300$ μV. From this value, taking into account the calibrated conversion between wave packet electron number and voltage response, we evaluate that we can unambiguously distinguish a wave packet of less than 100 electrons, equivalent to a $\pi$ phase shift, $90 \pm 5$ electrons within the interaction time ($\approx 100$ ps). For the results presented in this work and extrapolating the detection efficiency to the GHz regime for state-of-the-art electrometers in semiconductor nanostructures, we obtain comparable charge sensitivity (1 electron in $\approx 100$ ns).

## VI. FEW FLYING ELECTRON DETECTION

After calibrating the sensing system, we test it directly on electrons propagating in the ECs. We engineer a triggered flying electron wave packet in exciting the gate RF3 [cf. Fig. 1(a)] with a voltage pulse through capacitive coupling [40]. The generation pulse excites two EMPs, one at the rising edge and one at the falling edge [40,41], as sketched in Fig. 8. Applying such pulse with negative amplitude, the first front (falling edge) will excite electrons from the Fermi sea and hence induce a local increase of the ECs electron density. Such electron density variation is expected to induce an acceleration of the qubit oscillations which leads to a decrease of $\Delta V_{\text{RF}}$. On the contrary, the second front of the pulse (rising edge) induces a local decrease of the ECs electron density (cf. sketch in Fig. 8) and leads to an increase of $\Delta V_{\text{RF}}$. The number of flying electrons forming these EMPs is controlled by the generation pulse amplitude. To probe them we tune the spin qubit detector in its more sensitive configuration, $\varepsilon = \varepsilon_{3\pi/2}$ (7.5 mV in Fig. 7), set the generation pulse duration to 6 ns (see Fig. 9), and control the delay between the generation and detection pulses.

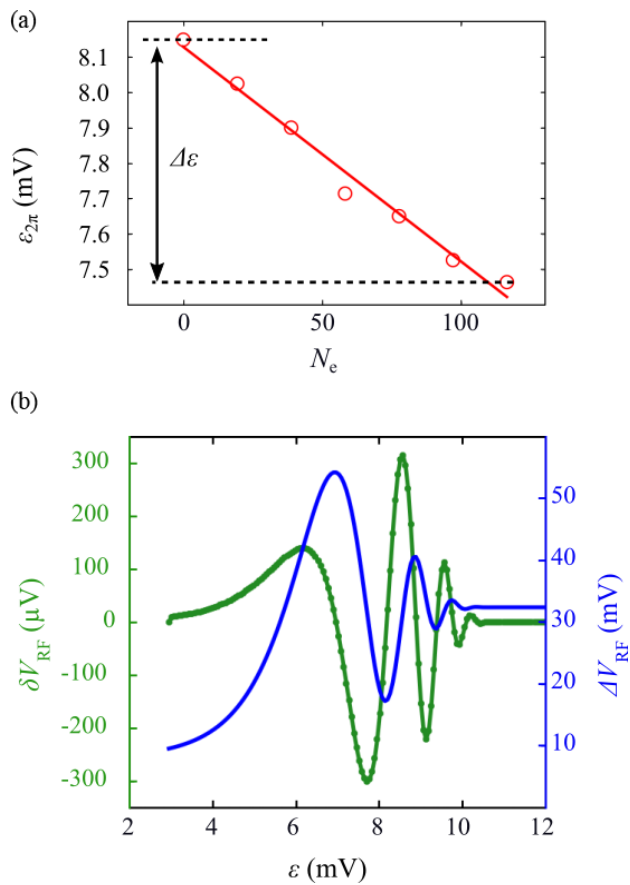

FIG. 7. (a) Detuning $\varepsilon_{2\pi}$ for which a $2\pi$ spin rotation of the ns-pulse coherent oscillations is achieved [see Fig. 6(b)], as a function of the number of electrons $N_e$ interacting with the spin qubit detector when applying a DC bias to the ECs. The $\varepsilon_{2\pi}$ values are extracted from the fit of the data shown in Fig. 6. From the linear fit of the data points (red line) we obtain a detuning shift of 6 $\mu$V per electron. (b) Estimated amplitude of the single-electron signal variation $\delta V_{RF}$ (green) and the voltage amplitude $\Delta V_{RF}$ of the spin triplet RF-QPC signal [blue: fit of data shown in upper Fig. 6(a)] are plotted against the detuning $\varepsilon$. The maximum signal is obtained for the $3\pi/2$ and $5\pi/2$ spin rotation positions, i.e., 7.5 mV and 8.5 mV.

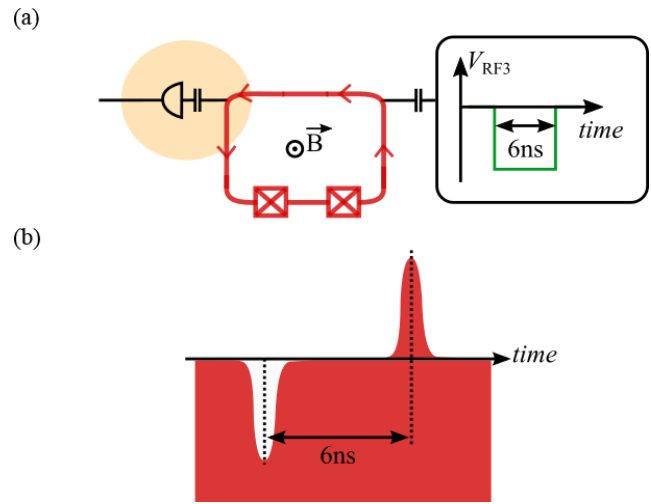

FIG. 8. (a) Sketch illustrating the coupling of the 6 ns generation pulse applied to a RF gate on the right part of the sample with the edge channels represented with the red line themselves capacitively coupled to the spin qubit detector. (b) This pulse generates two edge magnetoplasmons, one for each front of the pulse as sketched here with the electron density in red. The first front of the pulse being a negative voltage variation leads to an increase of the electron density, the second front, a positive voltage variation, a decrease. The two edge magnetoplasmons are separated in time by the 6 ns width of the generation pulse.

The flying electrons detection is first performed with a positive magnetic field of 430 mT such that the electrons travel anticlockwise and pass in the vicinity of the spin qubit detector (see Fig. 8). The induced signal variation, labeled $\Delta$, is shown in Fig. 10(a). We clearly observe two main peaks of opposite amplitudes separated by 6 ns for delays of $-3.5$ and 2.5 ns. The main peak full width at half maximum, of about 1 ns, is in agreement with the rise time of the voltage pulses. When reversing the magnetic field, the response peaks are completely suppressed [see Fig. 10(b)]. In this case, the EMPs are following the lower path of the ECs and are absorbed in the ohmic contact without passing next to the spin qubit detector. These elements strongly indicate that local detection on-the-fly of electrons has been implemented.

From the data presented in Fig. 10(a), the smallest number of flying electrons detected by repeating the experiment is obtained. Indeed, the noise level is of the order of $\sim$1 mV after averaging and limited by the slow drift of the qubit detuning. Taking into account the calibration (300 $\mu$V/electron), the procedure permits us to detect the passage of approximately 4 flying electrons.

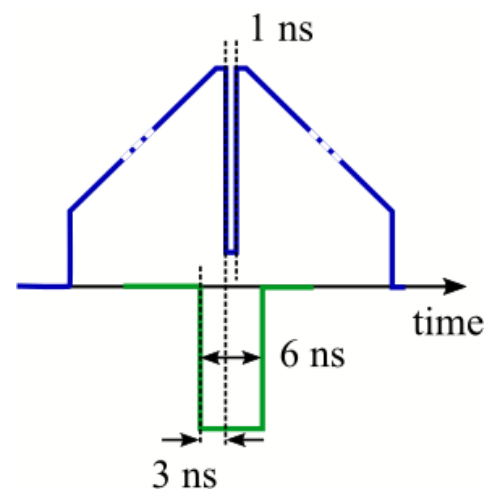

FIG. 9. Sketch of the pulse sequence to detect EMPs. In blue is the sequence applied on the spin qubit detector, in green the EMP generation pulse. By default the 1 ns detection pulse is centered on the generation pulse of 6 ns. Then, when sweeping it (in a 12 ns window with our setup) the detection can be synchronized with the rising (falling) edge of the generation pulse for positive (negative) delay about $\pm$3 ns.

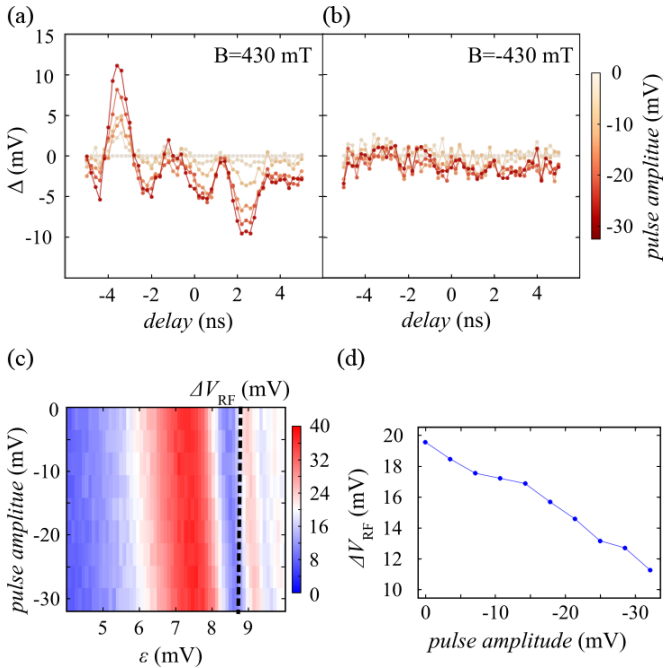

FIG. 10. (a) Detector signal $\Delta$ as a function of the time delay between the generation pulse and the detection pulse. $\Delta$ is the voltage amplitude $\Delta V_{RF}$ of the spin triplet RF-SQD signal averaged over 65 000 single-shot measurements from which the reference signal measured without generation pulse has been subtracted. This measurement is performed at a magnetic field of 430 mT, for which the flying electrons travel anticlockwise and pass nearby the spin qubit detector. The generation pulse amplitude, estimated on the sample, is represented in color scale. The signal $\Delta$ shows two main peaks for delays of 2.5 ns and −3.5 ns, of opposite sign and with an amplitude linearly increasing with the generation pulse amplitude. Such behavior tends to indicate the detection of edge magnetoplasmons. (b) Same experiment as (a) but with $B = -430$ mT, so opposite propagation direction in the edge channels. There is no signal variation, demonstrating the detection of flying electrons in the trace of (a). (c) Voltage amplitude $\Delta V_{RF}$ as a function of $\varepsilon$ for different voltage amplitude on the RF gate RF3 (cf. Fig. 1). The delay set to 2.5 ns means a detection synchronized with the falling edge of the generation pulse, so a decrease of the ECs electron density, for which the observed deceleration of the oscillations was expected. We can estimate the gate RF3 lever arm using the data presented in this paper. From panel (a), the maximum 32 mV of pulse amplitude induces a $\pi/4$ rotation. It is equivalent to about 0.3 mV of detuning shift. This phase variation is obtained for a DC bias close to 150 $\mu$V as shown in Fig. 5(b). From here we can estimate a lever arm on the ECs of about 1/200. This value is compatible with the geometry of the sample and the reported ones in quantum dots. It indicates that a few tens of flying electrons are generated for the maximum amplitude which agrees with the experimental data. (d) Cut at the $\varepsilon = \varepsilon_{3\pi/2}$ position of (c) corresponding to the black dashed line. The voltage amplitude $\Delta V_{RF}$ is decreasing when increasing the generation pulse amplitude, with a signal variation equivalent to the one shown in (a).

In addition to the main peaks, we observe secondary peaks at different time delays. Their origin is presently unclear. We speculate that they could arise from charge fractionalization due to the presence of several edge channel ECs [40–42].

To complement our study of the influence of the EMPs on the spin qubit, we performed an experiment mimicking the one shown in Fig. 6 but using flying electrons instead of a DC current. Since the spin qubit exposure time of 200 ps is smaller than the 1 ns width of the EMPs [cf. Fig. 10(a)], the two experiments are very similar when synchronizing the detection with an EMP. We chose the falling edge of the generation pulse (decrease of EC electron density) with a delay of 2.5 ns. The experimental data are shown in Figs. 10(c) and 10(d) showing a deceleration of the oscillations when increasing the generation pulse amplitude. This response is equivalent to the one presented in Fig. 6 where a current was flowing in the ECs. Therefore, it is another argument in favor of the detection of propagating charges in the ECs. Besides, the slice along the approximate position, limited by the $\varepsilon$ resolution, $\varepsilon = \varepsilon_{3\pi/2}$ [Fig. 10(d)], shows the same $\approx 10$ mV signal variation as presented in Fig. 10(a) with an expected linear evolution. By comparing the detuning shift of Fig. 10(c) to the equivalent DC bias of the calibration presented in Fig. 5(b), we obtain a lever arm on the ECs estimated to be about 1/200 eV/V. Such a value indicates that the generation pulse excites a few tens of flying electrons for the maximum amplitude [using Fig. 6(a)] which is in agreement with the experimental data. All these elements put together strongly suggest the successful development of a detector able to perform in-flight detection of flying electrons.

In conclusion, we demonstrated that a singlet-triplet spin qubit can be employed as an ultrasensitive charge detector able to detect a few flying electrons. The flying electrons are excited by a nanosecond voltage pulse and are propagating as EMPs in the ECs of the quantum Hall effect. The spin qubit is triggered by a nanosecond exchange pulse in a regime highly sensitive to electrical perturbation. Signal from as low as 4 flying electrons is obtained in averaged measurements. This detection scheme is stroboscopic and imposes synchronization of the detection pulse with the propagating electrons with a microsecond duty cycle. Intrinsically, it could eventually limit the throughput of quantum electron optic experiments. However, this two-stage detection scheme has nevertheless the potential of single-shot detection which would represent a significant step forward for quantum electron optics experiment. To reach this goal, it would first be crucial to enhance the capacitive coupling between the detector and the flying electrons through further optimizing the design of the interaction region. Also, presently this coupling is limited by the spreading of the electron wave packet over the many edge channels due to the low magnetic field. A stronger coupling could be achieved by using flying qubits operating at low magnetic field [43]. Operating the spin qubit detector at higher magnetic field is a solution in the limit of staying below 1 T [44]. Second, manipulating the spin qubit with higher fidelities would be necessary to increase the overall fidelity of the readout process. This is possible by reducing the charge noise and improving the RF-QPC charge sensitivity, from optimizing the impedance matching of the tank circuit and the SQD location.

## ACKNOWLEDGMENTS

We acknowledge technical support from the "Poles" of the Institut Néel, in particular the NANOFAB team who helped with the sample realization, as well as P. Perrier and C. Hoarau. A.L. and A.D.W. gratefully acknowledge the support of DFG-TRR160, BMBF–Q.com-H 16KIS0109, and the DFH/UFA CDFA-05-06. C.B. and T.M. acknowledge financial support from the French National Agency (ANR) in the frame of its program ANR Fully Quantum Project No. ANR-16-CE30-0015-02. T.M. acknowledges financial support from ERC "QSPINMOTION". C.B. acknowledges funding from the European Union's H2020 research and innovation program under Grant Agreement No. 862683.